# Theory of a stable strong electrostatic double layer generated in a two-grid Franck-Hertz tube

**Peter Nicoletopoulos**

*Faculté des Sciences, Université Libre de Bruxelles, Brussels, Belgium*

E-mail: pnicolet@skynet.be and pnicolet@ulb.ac.be

**Abstract**

There is a classic alternative to the Franck-Hertz experiment designed to show more than a recurrence of the first excited state. Instead of being subjected to a rising potential between source and accelerating grid, electrons are now accelerated in less than one excitation mean-free-path by an extra grid, and then drift towards the second grid across a large equipotential region. In this arrangement one must face the difficulty that the space potential between the grids is strongly modified by space charges. A recent analysis of this experiment with mercury showed that there is a particular form of discharge that generates and sustains the ideal design *dynamically*. The inevitable variations of potential inside the intergrid volume are then confined within a narrow sheath, a free double layer, joining two field-free plasma regions. The position of the double layer can be controlled so as to optimize the experiment. In the present paper, those phenomena are studied theoretically in steady-state conditions. The essence of the method is to specify velocity distributions for electrons and ions and use them to solve the Poisson equation. *Vortex-type* functions are used for trapped particles and mono-energetic beams for free particles. The model explains the change of position of the double layer and provides a natural explanation of a threshold condition for its amplitude that governs the transition between two critical configurations, in accordance with experiment. Stability and minimum field-energy considerations lead to a unique solution in a four-dimensional parameter space where the Langmuir ratio is equal to one. The method is adaptable to other experiments on double layer formation under discharge conditions and correctly predicts the values of several plasma parameters.

# 1. Introduction

The main object of this paper is to model theoretically the formation of a steady state electric double layer within the volume of bounded discharge plasma.

This project is part of a longtime effort [1-3] to elucidate a well-known extension of the Franck-Hertz experiment with mercury, designed to show more than the recurrence of the lowest excited state. In that version, rather than being subjected to a rising potential between source and accelerating grid, electrons are accelerated in less than one excitation free-path by an extra grid, and allowed to travel towards the second grid across a supposedly field-free region.

A large amount of structure in the current-voltage curve is indeed displayed, but it is very difficult to find a setting of the energy scale leading to a consistent interpretation. The foremost problem is that the electrostatic potential distribution evoked by the electrode voltages is strongly modified inside the intergrid region by the interaction of negative and positive space charges.

A detailed analysis (see [2]) showed that a determination of the space potential can be deduced from a careful examination of the change in the spectra induced under controlled conditions. It was found that if a constant *decelerating* voltage ($\Delta V$) is applied across the main collision cell, the series of peaks is identifiable as a superposition of two copies of $e^-$-Hg spectra excited at two different potentials. The obvious inference is that the scattering cell is separated into two field-free, weakly ionized plasma regions, joined by a narrow space-charge double layer (see figure 1). Also required for a complete interpretation is the presence of another space-charge structure, a rising wall-sheath at the exit of the first grid, as shown by the dashed curve in figure 1.

Evidently, besides its main scope, this experiment is also a useful workground for studying the formation of narrow electrostatic structures in a bounded region of plasma containing neutral gas at moderate pressures. The advantage of this particular experimental arrangement is that the link to specific atomic physics phenomena stands out immediately.

Lacking as yet is a direct confirmation of the presence of a wall-sheath and a 'free' double layer (DL) during operation; nor is it likely that a non-intrusive measurement of the space potential is forthcoming in this type of sealed apparatus. Suggestive as it may be, the testimony presented so far to uphold the contention of DL formation is purely circumstantial.

But a more persuasive case would be made if one could put together a physical model that could be subjected to mathematical analysis to show the emergence of the proposed reconfiguration of the space potential for appropriate values of the plasma parameters. Above all, the theory should answer the crucial question: *How is it possible for this potential structure to persist throughout the experiment despite the ostensible alteration of the various components of space charge as the energy of the primary beam is raised?*

Section 2 is a collection of the experimental facts found in [2], supplemented by some additional observations, obtainable electrically and visually, on the evolution of the discharge towards the stable form governed by the space potential shown in figure 1. This sum-up clarifies the objectives of the model. In the interest of simplicity the physical arguments (and the subsequent theory) are presented in infinite parallel plane geometry. In the actual apparatus only the first grid is planar. The other three electrodes are cylindrical.

The remaining sections are devoted to formulating the theory and using it to model the observations of section 2.

## 2. Review and appraisal of observations

*2.1 Inferences gathered from examination of excitation curves*

The configuration shown in figure 1 sets in abruptly at a well-defined value, say $V_{crit}$, of the accelerating potential $V_{g_1}$. Past this point, the excitation curve (namely the plot of the current between $g_2$ and the anode A against the accelerating potential $V_{g_1}$) displays two copies of the $e^-$-Hg spectrum. One of them corresponds to inelastic events incurred at potential $V_{g_1}+\delta$ and is independent of the value of $\Delta V$. The other copy corresponds to the same series of inelastic events incurred at potential $V_{g_1}-\Delta V$, namely the voltage of grid 2. Increasing $\Delta V$ displaces the latter series of peaks rightwards with respect to former, and vice-versa. One can thus optimize the experiment by adjusting the value of $\Delta V$ to hide the high-potential peaks.

A typical 'pure' excitation curve obtained in these conditions is shown in figure 2. This curve was taken with $\Delta V$=4.1. The peaks labelled by numbers **2,3,4**... show inelastic events incurred in the low-potential plasma (their energies are measured from **D** at $V_{g_1}=\Delta V$). Peaks **b,c,d** correspond to events **2,3,4** incurred at $V_{g_1}$ potential, and are always present since the low-potential drift space has not been formed as yet. Peak **1** has no bearing on the present considerations; it is due to a shape resonance in the elastic cross section at 0.4 eV as discussed at length in [3].

The discontinuity labelled **A** (visible here as a precipitous completion of peak **2**) is a signal that the space potential on the high potential side jumps from the solid to the dashed line in figure 1. The position of feature **A** (namely $V_{crit}$) and the value of $\delta$ are affected by the strength of cathode current, but are independent of the value of $\Delta V$. At optimal cathode current (as in figure 2) feature **A** occurs at ~9.5 V on the scale of $V_{g_1}$ potential, and $\delta$=2.6 volts. Hence the total amplitude $\psi=\Delta V+\delta$ in figure 2 is 6.7 volts. At larger currents, the system goes into an oscillating regime at feature **A** (for details on this point see section 8.2 of [2]).

At values of $\Delta V$ lower than 4.1, the peaks of figure 2 are shifted leftwards and so the high-potential features are gradually unmasked (see figures 2, 3 and 4 of [2] where also feature **A** is more clearly displayed).

Below $\Delta V$=2.1 the low-potential peaks disappear and the curves become pure high-potential spectra (though peaks **2** and **3**, now lying well below feature **A**, may remain discernible). This leads to the most important inference: *a minimum intergrid voltage, $\psi=\Delta V+\delta=2.1+2.6=4.7$, is necessary for the sustained presence of a field-free low-potential region of significant width throughout the range of accelerating potential.* Moreover, *the energy equivalent of 4.7 V coincides with the lowest inelastic threshold of mercury.*

Values of $\Delta V$ between 4.1 and 4.3 offer the best display of the spectrum. For $\Delta V$>4.3, the high-potential copy does not reemerge but the peaks become significantly weaker.

*2.2 The role of cathode emission current*

The formation of a double layer well away from grid 2 is also crucially dependent on the strength of the current emitted by the oxide-coated cathode. Too low a cathode current prohibits the formation of low-potential plasma: regardless of the amplitude $\psi$ of intergrid potential only the high-potential spectrum is observable. Evidently the DL remains very close to the outer grid in this case.

Strong as the injected beam needs to be, it seems that it must also be sufficiently cold. Overheating a poorly activated cathode to raise the current will not do. *The requirement of a sufficiently strong and cold electron beam is a second important condition for moving the double layer away from grid 2.*

*2.3 Positive-ion current*

The ion current can be monitored separately by a small modification in the anode circuit. Curves recorded with that arrangement show that, with a well activated cathode, a weak ion current sets forth already at feature **b**, likely due to cumulative ionization via the metastable $^3P_0$ and $^3P_2$ states. At $V_{g_1}=V_{crit}$ (feature **A**) this ion current rises suddenly by several orders of magnitude and increases steadily but slowly thereafter.

*2.4 Visual observations*

It is possible to ascertain some highlights in the evolution of the discharge by peering through the top of the tube while tracing curves of the type shown in figure 2. As the voltage $V_{g_1}$ is raised, the visible part of the region between the electrodes appears at first dark. When $V_{g_1}$ reaches the potential of peak **b**, two localized blue-green glows are born, seemingly bordering the flat sides of the electron gun, and stretching in opposite directions towards the second grid as the voltage increases. In different samples of the tube the rates of elongation in each direction may differ. When the outer edge of a glow reaches $g_2$, the entire half-volume on that side is lit up and the color changes to pinkish. This transition occurs precisely *at the dip preceding feature 2*, (feature **L** in figure 2). Thereafter the light becomes whiter and more brilliant.

The ring between $g_2$ and the anode remains dark. The persistent absence of visible light in that zone indicates that most inelastic collisions occur outside the $g_2$-A region (this is fortunate for otherwise the superior detection mechanism responsible for displaying the low-potential spectrum, as modelled in [3], would not be so simple).

On the scale of $g_2$ potential, **L** is located at 4.7-4.8 V. It is manifest that a special discharge to the outer grid is initiated when the beam energy at $g_2$ exceeds the level of the first excited state, $6^3P_0$, (a metastable state). As a result, a sufficient extent of field-free plasma adjacent to $g_2$ is immediately created with resulting emergence of peak **2** due to the next state ($6^3P_1$).

*2.5 A scenario of potential formation*

With the aid of these observations we can attempt to construct a plausible scenario leading to a steady space potential of the type shown in figure 1.

1) At features **b,c** ($V_{g_1} \cong 5$-5.5) two-step ionization via the metastable $^3P$ states (especially $^3P_2$) produces an excess of positive space charge. This creates a maximum of potential on either side of $g_1$ wherein the primary electrons are able to excite higher-lying states. For the green and blue lines of mercury to emerge the $7^3S_1$ state at 7.73 eV must be excited. The peaks of the humps must then be about 2.3-2.8 V above the voltage $V_{g_1}$. Ion production is maintained because the metastable states are being persistently repopulated by the radiative transitions $7^3S_1 \rightarrow ^3P_2$ (green) and $7^3S_1 \rightarrow ^3P_0$ (violet). As the electrode voltages $V_{g_1}$ and $V_{g_1}-\Delta V$ increase the outer parts of the space potential are raised and so the luminous regions stretch towards $g_2$. Electrons drift across the tube in opposite directions along narrow bunches of field lines emanating from the planar exits of the electron gun.

This stage of the discharge is reminiscent of the physical picture based on volume ionization modelled in Langmuir's seminal paper, where the space potential turns out to have a maximum between cathode and anode in a parabolic configuration (see [4] pp 961-963). The temporary lack of uniformity of the uphill potential is in line with the distortion of peaks **a,b,c** (as compared to their analogs **2,3,4**).

2) At $V_{g_1} \cong 4.7+\Delta V$ the expanding glows reach $g_2$. Somehow, the electron and ion populations now conspire to give rise to a region of field-free plasma adjacent to $g_2$ (this is *'somehow number 1'*). In this space the transport of charge carriers becomes essentially diffusive. Hence ions and electrons spread out transversely to the field lines and so the glow fills the entire intergrid space.

3) The stage has now been set where a short further increase of $V_{g_1}$ will lead to the configuration along the dashed line of figure 1. Close to feature **A** the energy of the primary beam at the hump of space-potential reaches the ionization threshold (10.4 eV), the ion current surges, and (again somehow) the hump expands into an equipotential plateau at potential $V_{g_1}+\delta$ (this is *'somehow number 2'*). The uphill region has now confined the previous smooth variations of electric field into a narrow sheath hugging $g_1$. From then on peaks corresponding to excitations at $V_{g_1}+\delta$ potential (now undistorted) will be displayed. As mentioned, the maximal value of $\delta$ is always close to 2.6 volts.

The strife among the various components of space charge to establish overall stress balance leads to a pattern of two plasmas separated by a narrow DL. The continued presence of wide enough low-potential plasma is critically dependent on the magnitude of $\Delta V$. Unless $\psi$ exceeds the inelastic threshold of mercury *and* the primary beam is sufficiently strong and cold, the high-potential plasma will ultimately invade most of the intergrid region.

*2.6 Goals of the theory*

Our object is to construct a steady state model applicable to the final stage. Unlike existing theories, this one should be capable of producing *unequal* widths of low and high-potential plasma and should thus reveal the mechanism leading to the two extreme configurations where one or the other of the two plasmas invades most of the intergrid region.

Hopefully, the model will elucidate the basic requirements of a critical amplitude $\psi$ and a sufficiently strong and cold primary electron beam and provide valuable insight in regard to the *somehows* of subsection 2.5.

The fast transitions at **L** and **A** require a time dependent approach and will not be treated in this paper.

## 3. Formulation of the model

Our approach is similar in spirit to the one-dimensional kinetic model for steady stade monotonic double layers of the 'strong' variety (namely of unlimited amplitude) introduced by Schamel [5-6].

The basic ingredients consist of four species of particles: free and trapped electrons and ions. Their velocity distributions are functions of the particles' total energy. The free (drifting) particle distributions $f_{ef}$ and $f_{it}$ are represented by delta functions and the trapped ones $f_{et}$ and $f_{it}$ are a generalized sort of truncated Maxwellians:

$$f_{ef} = n_0 [v_0/(v_0^2 + 2\phi)^{1/2}] \delta[v - (v_0^2 + 2\phi)^{1/2}] \qquad (1a)$$

$$f_{et} = A \exp[(\beta/2)(v^2 - 2\phi)] \qquad v^2 < 2\phi \qquad (1b)$$

$$f_{if} = \nu n_0 [u_0/[u_0^2 + 2(\psi-\phi)]^{1/2}] \delta\{u + [u_0^2 + 2(\psi-\phi)]^{1/2}\} \qquad (1c)$$

$$f_{it} = \nu B \exp\{(\alpha/2)[u^2 - 2(\psi-\phi)]\} \qquad u^2 < 2(\psi-\phi) \qquad (1d)$$

Electron velocities $v$, $v_0$ and ion velocities $u$, $u_0$ are normalized by $(T/m)^{1/2}$, $(T/M)^{1/2}$, respectively, and potentials $\phi$, $\psi$ by $T/e$. M and m are the ion and electron masses, e is the electron charge, T is an arbitrary thermal energy, $n_0$ is the free-electron density at $\phi=0$, and A, B, $\nu$ are positive constants.

The parameters $\alpha$ and $\beta$ can be of either sign. Negative values correspond to common (truncated) Maxwellians. Positive values generate concave (vortex type) functions. The extension to positive $\alpha$, $\beta$ was introduced by Schamel (see [5-6]) in order to model the formation of 'electron and ion holes' and of so-called 'slow particle acoustic' DL's. The limits $\alpha \to 0$, $\beta \to 0$ correspond to rectangular ('water-bag') distributions.

The assumption that an unisotropic component with a single narrow peak is able to capture the essential physics might seem questionable, but will be justified later on. Note that the direction of the beams can be changed by inverting the sign of the variables v and u. This sign is immaterial however, because the subsequent results depend only on $v^2$ and $u^2$.

The fact that our distributions are functions of the total energy (thus satisfying the Vlasov equation) does not imply that our model is collisionless. It simply means that the length d of the discharge is not significantly larger than the mean free path $\lambda_\varepsilon$ for energy transfer to neutrals [$\lambda_\varepsilon = \lambda_M (M/2m)^{1/2}$ where $\lambda_M$ is the momentum-transfer cross section]. This is indeed true in the pressure range (3-4 Torr) of this experiment although a large number of momentum-changing collisions actually take place (see [3]).

In a first attempt [7] to model the phenomena of the extended Franck-Hertz experiment the distributions (1b) and (1d) were used with negative $\alpha$ and $\beta$. Later work revealed that strong DL solutions are also possible for $\alpha, \beta > 0$ and that above all the latter case is favored by experiment (see section 11).

The electron and ion densities obtained from (1a)-(1d), normalized by $n_0$, are given (for positive $\alpha$ and $\beta$) by

$$n_e = a\exp(-\beta\phi)\operatorname{erfi}(\beta\phi)^{1/2} + \left[1 + \frac{(X^2-1)\phi}{\psi}\right]^{-1/2} \quad (2a)$$

$$n_i = b\nu\exp[-\alpha(\psi-\phi)]\operatorname{erfi}[\alpha(\psi-\phi)]^{1/2} + \nu\left[1 + \frac{(Z^2-1)(\psi-\phi)}{\psi}\right]^{-1/2} \quad (2b)$$

where $X = (1 + 2\psi/v_0^2)^{1/2}$, $Z = (1 + 2\psi/u_0^2)^{1/2}$, and erfi(z) is the 'imaginary error function' given by erfi(z)=erf(iz)/i. The function $\exp(-\beta\phi)\operatorname{erfi}(\beta\phi)^{1/2}$ (the 'Dawson integral') vanishes at $\phi=0$ and so there are no trapped electrons at the low-potential boundary (and no trapped ions at the high-potential boundary). The constants a and b measure the proportion of trapped particles and $\nu$ is the ratio of the high and low-potential densities.

Self-consistent solutions for the space potential are sought by solving the Poisson equation

$$\frac{d^2\phi}{dx^2} = n_e - n_i \equiv -\frac{dV(\phi)}{d\phi} \quad (3)$$

where $V(\phi)$ is the 'Sagdeev' (or 'classical') 'potential' and x is the space coordinate normalized by the electron Debye length $\lambda_D$, expressed in cm by

$$\lambda_D = 743\left(\frac{T}{n_0}\right)^{1/2} \qquad (4)$$

with T in eV and $n_0$ in cm$^{-3}$. From now on we set T=1 so that potentials are expressed directly in volts.

Integrating equation (3) once, we get

$$\frac{1}{2}\left(\frac{d\phi}{dx}\right)^2 + V(\phi) = 0 \qquad (5)$$

where the integration constant is chosen such that the field $E(x)=-d\phi(x)/dx$ vanishes at $\phi=0$, namely $V(0)=0$. $V(\phi)$ is thus found to be

$$V(\phi) = \frac{2\psi}{X^2-1}\left[1-\left(1+\frac{(X^2-1)\phi}{\psi}\right)^{1/2}\right] - \frac{a}{\beta}\left[2\left(\frac{\beta\phi}{\pi}\right)^{1/2} - \exp(-\beta\phi)\text{erfi}(\beta\phi)^{1/2}\right]$$
$$+ \frac{2\nu\psi}{Z^2-1}\left[Z-\left(1+\frac{(Z^2-1)(\psi-\phi)}{\psi}\right)^{1/2}\right] - \frac{b\nu}{\alpha}\left[-2\left(\frac{\alpha\psi}{\pi}\right)^{1/2} + 2\left(\frac{\alpha(\psi-\phi)}{\pi}\right)^{1/2}\right] \qquad (6)$$
$$- \frac{b\nu}{\alpha}\left[\exp(-\alpha\psi)\text{erfi}(\alpha\psi)^{1/2} - \exp[-\alpha(\psi-\phi)]\text{erfi}[\alpha(\psi-\phi)]^{1/2}\right]$$

There are three further boundary conditions to be satisfied

$$V(\psi)=0 \qquad (7)$$

$$n_e(0)-n_i(0)=0 \qquad (8)$$

$$n_e(\psi)-n_i(\psi)=0 \qquad (9)$$

Equations (8) and (9) can be used to express a and b in terms of $\nu$

$$a = \frac{(X\nu-1)\exp(\beta\psi)}{X\text{erfi}\sqrt{\beta\psi}} \qquad (10)$$

$$b = \frac{(\nu-Z)\exp(\alpha\psi)}{\nu Z\text{erfi}\sqrt{\alpha\psi}} \qquad (11)$$

$a(\nu)$ and $b(\nu)$ are then substituted in (7) which is solved for $\nu$, whence

$$\nu = \frac{\dfrac{1}{\alpha} + \dfrac{1}{\beta X} + \dfrac{2\psi}{1+X} - \dfrac{2\sqrt{\psi}\exp(\alpha\psi)}{\sqrt{\pi\alpha}\,\text{Erfi}\sqrt{\alpha\psi}} - \dfrac{2\sqrt{\psi}\exp(\beta\psi)}{X\sqrt{\pi\beta}\,\text{Erfi}\sqrt{\beta\psi}}}{\dfrac{1}{\beta} + \dfrac{1}{\alpha Z} + \dfrac{2\psi}{1+Z} - \dfrac{2\sqrt{\psi}\exp(\alpha\psi)}{Z\sqrt{\pi\alpha}\,\text{Erfi}\sqrt{-\alpha\psi}} - \dfrac{2\sqrt{\psi}\exp(\beta\psi)}{\sqrt{\pi\beta}\,\text{Erfi}\sqrt{\beta\psi}}} \tag{12}$$

Using (10)-(12) one obtains $V(\phi)$ as a function of the five remaining parameters $\psi$, $\alpha$, $\beta$, $X$ and $Z$. The results to be described below were obtained by trial and error from various plots of this final form of $V(\phi)$.

## 4. Three important properties of the model

Consider first the symmetric case $\alpha=\beta$, $Z=X$ (and hence $\nu=1$). Figure 3(a) shows the function $V(\phi)$ obtained for $\psi=3$, $\alpha=0.5$, $X=1.1$. This has the familiar inverted bell-type form that is traditionally taken as proof of the desired configuration: a sheath of rising potential joining two unbounded quasi-neutral and field-free regions, ideal plasmas. Unbounded, because in all previous models the mathematical expression for the total distance

$$\Lambda(\psi) = x(\psi) - x(0) = \int_0^\psi \frac{d\phi}{\sqrt{-2V(\phi)}} \tag{13}$$

diverges (logarithmically) at the end points $\phi=0$ and $\phi=\psi$: $V(\phi)\sim\phi^2$ and $\sim(\psi-\phi)^2$ as $\phi$, $\psi-\phi$, $\to 0$, since imposition of ideal plasma conditions at the boundaries removes the zeroth and first powers.

The first important point in the present model is that $\Lambda$ is *finite*: an uncancelled half-power remains so that $V(\phi)\sim\phi^{3/2}$ and $\sim(\psi-\phi)^{3/2}$ at the limits: truncating the trapped distributions and removing the high-energy tails of the free distributions produces natural cutoffs at the previously singular endpoints of the integral in (13).

The second major point is that $V(\phi)$ may develop *internal* zeros: As $X$ is increased, the edges of the bell fold upwards as in figure 3(b). Eventually the folds turn into local maxima, and gradually rise toward the $\phi$ axis, to become 'quasi-zeros' of $V(\phi)$ at points $\phi=\rho$ and $\phi=\psi-\rho$ as shown in figure 3(c). The quasi-zeros are found to be quadratic, namely

$$-V(\rho)\sim(\phi-\rho)^2+\varepsilon, \qquad -V(\psi-\rho)\sim[\psi-\rho-\phi]^2+\varepsilon \qquad 0<\varepsilon\ll 1 \tag{14}$$

As X is further increased the maxima creep into the region of positive $V(\phi)$ ($\varepsilon<0$) and the solutions become oscillatory. Such solutions are not discussed in the present paper.

The quasi-zeros give rise to narrow peaks in the integrand of (14) and hence to extended quasi-field-free regions in x-space. In this way we obtain a quantitative description of plasma and double layer formation in a bounded region of $\Lambda$ Debye lengths.

Also present are 'pre-sheaths' of amplitude $\sim\rho$ at the boundaries (see figure 5(a) in section 6). Note that the Bohm condition for the sheath edge at $\phi\sim\rho$, expressed (as shown in [8-9]) in terms of $V(\phi)$, by $d^2V/d\phi^2<0$ at $\phi=\rho$, is equivalent to requiring the generation of quasi-zeros.

The quantity $\Lambda(\psi)$ increases or decreases (logarithmically) as X is 'tuned' or 'de-tuned' to lower or raise the value of $\varepsilon$. As long as $\varepsilon$ remains small, this stretches or contracts the plasma regions without affecting the pre-sheaths and the double layer.

The third important property is that if $\psi$ is varied, with $\alpha\psi$, $\beta\psi$, X and Z fixed, $V(\phi)$ remains self-similar: the function $(1/\sigma)V[\sigma\phi, \sigma\psi; \alpha/\sigma, \beta/\sigma, X, Z]$ is independent of $\sigma$.

Decreasing the product $\alpha\psi$ requires a smaller value of X for obtaining quasi-zeros and moves them closer together (and vice versa). Below a certain value ($\alpha\psi\cong1.05$), the quasi-zeros coalesce [$-V(\rho)\sim(\phi-\rho)^4+\varepsilon$] so that only a single plasma region is left between large pre-sheaths of amplitude $\psi/2$. For large $\alpha\psi$ the amplitude of the pre-sheaths can be very small so that we have essentially two plasma regions at 'wall' potential.

The formation of internal quasi-zeros of $V(\phi)$ is not particular to the present theory. Langmuir's introduction of the term 'plasma' for the field-free region following his 'double-sheath' ([4] pp. 976-980) was founded on exactly this mechanism (in a simpler model with one-sided boundary conditions). Latter-day authors have ignored this feature although it probably occurs in any model constructed with trapped and drifting species. A typical example is the kinetic model proposed by Swift [10]; it is easily verified that his Sagdeev potential too can acquire one internal quasi-zero (but never two) for certain values of the parameters.

Needless to say, the imposition of exact charge neutrality at the boundaries is no longer physically necessary. Relaxing the boundary conditions (8) and (9) to some extent does not significantly change the form of V($\phi$) and its quasi-zeros, but would require more parameters. Thus, setting n(0) and n($\psi$) strictly equal to zero is merely a matter of mathematical convenience.

An interesting consequence of the finiteness of $\Lambda$ for non-zero $\varepsilon$ is that equation (13) provides a determination of the quantity $n_0$, the density of free electrons at the low-potential boundary, in a plane-parallel discharge of length d cm and amplitude $\psi$ volts (T=1):

$$n_0 = 743^2 [\Lambda(\psi)/d]^2 \text{ cm}^{-3} \qquad (15)$$

## 5. Formation of two plasma regions of unequal length

For unequal values of $\alpha$, $\beta$, separated quasi-zeros are generated regardless of the smallness of $\alpha\psi$, $\beta\psi$.

For $\beta\psi < \alpha\psi$, we find that

1) The low and high-potential quasi-zeros ($\rho_{low}$, $\psi$-$\rho_{high}$) satisfy $\rho_{low} > \psi$-$\rho_{high}$.
2) The higher quasi-zero produces a peak of smaller area in the integrand of (13). Therefore, the low-potential field-free region is widest. In other words the DL is closer to the high-potential boundary.
3) $\nu > 1$.

For $\beta\psi > \alpha\psi$ the converse statements are true.

## 6. A minimum energy principle

From the preceding discussion one is inclined to take the view that, in the types of discharge considered here, only the solutions containing quasi-zeros are physically relevant. In other words, around a region of parameter space that produces admissible (bell-type) forms of V($\phi$) the system always strives to adjust to the closest allowable configuration with two quasi-zeros.

To justify this pronouncement we need some physical principle, like a minimum of some energy. The obvious quantity is the total electrostatic field energy F of the space charge configuration between $\phi=0$ and $\phi=\psi$. This is easily evaluated in terms of $V(\phi)$

$$F = \int_{x_0}^{x_\psi} \left(\frac{d\phi}{dx}\right)^2 dx = \int_0^\psi \left(\frac{d\phi}{dx}\right)^2 \frac{dx}{d\phi} d\phi = \int_0^\psi \sqrt{-2V(\phi)}\, d\phi \qquad (16)$$

Figure 4 shows F in the symmetric case of section 4 ($\psi=3$, $\alpha=\beta=0.5$, X=Z), plotted for X between 1.03 and 6.675. We see that the field energy falls sharply at both ends of this range. Those limits of minimal energy obviously correspond to configurations with the largest extent of field-free space (and hence proportionally the narrowest double layer).

The minimum of the left branch develops at X~1, namely when the incident drift velocities are large, $v_0^2 \gg \psi$. Fairly field-free regions of some extent are formed as X→1, Z→1, because the leading terms of $V(\phi)$ near the boundaries have the form $V(\phi) \sim (1-X)\phi^{3/2}$, $\sim (1-Z)(\psi-\phi)^{3/2}$.

This is a far less efficient mechanism for generating field-free space than the alternative way based on internal quasi-zeros (see figure 5). Moreover, the fast-beam solutions do not comply with actual conditions in the bulk of a discharge: the primary beam is dissipated by inelastic impacts with resultant creation of substantial populations of slow charge carriers.

Clearly, it is the lowest field-energy limit (at $v_0^2 \ll \psi$, $u_0^2 \ll \psi$) that is consistent with our physical picture. The very interpretation of the spectrum in figure 2 rests on the premise that slow electrons are emitted isotropically from a source at the entrance of the $g_2$-anode region (see [3]).

It is conceivable that those electrons acquire the identity of a narrow beam, because a swarm of electrons driven through mercury gas by a small electric field, attains a velocity distribution with a narrow peak centered at nearly zero energy [11]. This is a consequence of a large peak at 0.4 eV in the momentum transfer cross section, due to a resonance in the e⁻-Hg system [3].

Analogous circumstances should hold for the ions, the slowing mechanism now being charge-transfer. The cross section for charge-transfer collisions of mercury ions in their own gas is similarly peaked at very low energy.

Figure 6 shows how the distance $\Lambda$ varies in the range of X of figure 4. We see that the limits of minimal energy correspond to the largest values of $\Lambda$. It follows from equation (16) that in discharges of this type filling a length d, the states of minimal energy are those of smallest Debye length (most efficient screening) and hence of largest charge density $n_0$.

## 7. A stability argument

Examining the evolution of the density ratio $\nu$ [equation (13)] can further focus the search for the most appropriate region in parameter space.

Plotting $\nu$ versus $\psi$ for various values of the set $\{\alpha, \beta, X, Z\}$ brings out some interesting features. Firstly, the form of $\nu(\psi)$ is essentially governed by the parameters $\alpha$, $\beta$, rather than X, Y. Secondly, for $\beta<\alpha$, $\nu(\psi)$ contains a maximum (a minimum for $\beta>\alpha$); the rate of rise is much more rapid than the rate of fall and the peak is rather blunt. Reducing $\beta$ shifts the position of the maximum rightwards while increasing $\alpha$ shifts it leftwards.

This indicates that there is a range of stability and suggests that we should seek values of $\alpha$, $\beta$ which place the maximum near the experimental value of $\psi$. In that way, small changes of amplitude in either direction would not require significant changes at the boundaries.

For moderately small $\beta\psi$ and moderately large $\alpha\psi$ the maximum of $\nu$ remains conspicuous and the DL is closer to but still free of the high-potential boundary. This is the range that we shall use below (section 8.1) in modelling the case of wider low-potential plasma between B and $g_2$ in figure 1.

In the extremely asymmetric situation where $\beta\psi\ll1$ and $\alpha\psi\gg1$ the rate of descent of $\nu(\psi)$ beyond the maximum becomes very weak : essentially, the function $\nu(\psi)$ rises exponentially to a constant. In this domain the DL reduces to a 'wall' double-sheath at the high-potential boundary. These solutions are remarkably robust. Provided only that Z is larger than about 20, the value of X required for producing the lower quasi-zero is unique: $X\cong9$. Apart from that these solutions are independent of $\alpha$, $\beta$, $\psi$ and the density ratio tends to $\nu \cong 2.8$ (see the Appendix). Surprisingly, this value of $\nu$ is a universal upper limit. This is an interesting result that can be tested experimentally. The largest value of $\nu$ that I have found in the literature is 2.5 [12]; but then this was in the free-DL domain where the theoretical values of $\nu$ are smaller than 2.8.

The stability of double layers is commonly discussed in the literature in terms of the 'Langmuir ratio'

$$L = \frac{v_0}{\nu u_0} = \frac{\sqrt{Z^2-1}}{\nu\sqrt{X^2-1}} \qquad (17)$$

It has been argued that the stablest double layers are those where L=1 (see [8-9] and references therein). Obviously in the present case $L(\psi)$ develops a minimum at the maximum of $\nu(\psi)$ (or a maximum for $\beta > \alpha$). It turns out that the value of $L(\psi)$ at the extremum is indeed close to 1 in moderately asymmetric circumstances (moderately small $\beta\psi$ and moderately large $\alpha\psi$) and changes somewhat in more extreme cases.

The outcrop of this analysis is that, given values $\psi_0$ and $L_0$, the constraints that (a) the extremum of $\nu(\psi)$ occurs at $\psi_0$ and (b) there are two quasi zeros, determine the set $\{\alpha, \beta, X, Z\}$ uniquely.

## 8. Calculating the space potential of figure 1

*8.1 The case of a larger proportion of low-potential plasma ($\beta < \alpha$)*

Step 1: The region between B and the second grid.

We take the high-potential boundary to be at the 'edge' B of the sheath formed at the exit of grid 1. To be precise, B is a point left of which there are no trapped electrons. The low-potential boundary is in a plane near grid 2 (see figure 1). Note that although the imposition of equation (8) at $g_2$ is still a matter of convenience, charge neutrality and vanishing electric field must both be enforced at B.

The observations outlined in section 2 suggest that we look for solutions in the range $4.7 < \psi < 6.7$. These solutions should display widest low-potential plasma, hence $\beta < \alpha$. We also suppose that L=1. We begin by choosing a value of $\psi$, say $\psi=5$, seek $\alpha$ and $\beta$ that place $v_{max}$ near $\psi=5$, and by trial, adjust X and Z to ensure that $V(\phi)$ has two good quasi-zeros ($\varepsilon$ close to zero) while L is kept close to 1. A good result is obtained with the following values for the basic parameters

$$\psi=5 \quad \beta=0.17 \quad \alpha=1.48 \quad X=10.695 \quad Z=21.338$$

for which the values of $v_{max}$ and of the Langmuir ratio L are

$$v_{max}=1.961 \text{ (at } \psi=4.995) \quad L=1.02$$

The quasi zeros, $\rho_{low}$, $\psi-\rho_{high}$, are at 0.15 and 0.037 respectively.

In view of the property of self-similarity of section 4, the parameters $\beta\psi=5\times0.17=0.85$, $\alpha\psi=5\times1.48=7.4$, with the above values of X and Z, define a unique solution characterized by the above values of $v_{max}$ and L. Larger values of $\psi$ imply proportionally smaller trapping parameters and larger incoming drift velocities. Stability is maintained because the maximum of $v(\psi)$ is always at the imposed value (say $\psi_0$) of $\psi$. The functions $v(\psi)$ corresponding to $\psi_0=5$ and $\psi_0=6.7$ are shown in figure 7a.

Figure 7b shows the space potential obtained from this global solution for $\psi=6.7$ (the experimental value used in recording figure 2). As required, the low-potential region is considerably wider.

It is conceivable that the discharge is able to adjust the set $\{\alpha, \beta, v_0, u_0\}$ so that the same solution is applicable to an entire interval of $\psi$, say $4.7<\psi<6.7$. However, as $\psi$ is raised to $\sigma\psi$, the distance $\Lambda$ is changed to $\sigma^{1/2}\Lambda$. Thus if the total length of the discharge is constant, the system must also adjust $n_0$ to $\sigma n_0$, with $\sigma$ as large as 1.4. An alternative is to imagine that X and Y are slightly de-tuned in the course of the adjustment process to increase the value of $\varepsilon$. This would keep $\Lambda$ unchanged and would not perceptively change the position of the maximum of $v$ in relation to the chosen value of $\psi$.

The preceding treatment can easily be generalized by adding a fast electron beam (say of initial velocity $w_0 \gg v_0$, such that $Y=(1+2\psi/w_0^2)^{1/2}$ is smaller than about 3), or a second slow beam whose velocity is comparable to $v_0$. On doing this one finds that the results are virtually unchanged: a single slow beam fully represents the essential features.

Step 2: The region between the first grid and B

The requirement of no trapped electrons at the left of point B is implemented by simply setting a=0 in equation (2a). Charge neutrality at $\phi=\psi$ [equation (9)] is mandatory and so is equation (7). The former condition implies

$$\nu = X^{-1} \qquad (18)$$

Inserting (18) in (7) and solving for b we obtain

$$b = \frac{2\psi\alpha(\frac{1}{1+Z} - \frac{X}{1+X})}{2\sqrt{\frac{\alpha\psi}{\pi}} - \exp(-\alpha\psi)\mathrm{erfi}\sqrt{\alpha\psi}} \qquad (19)$$

Using (18), (19) with a=0 in (6) we find our new $V(\phi)$ in terms of the four parameters $\psi$, $\alpha$, X, Z.

Solutions that also satisfy equation (8) do not exist, so that charge neutrality at grid 1 is impossible. Moreover, without trapped electrons there can never be a low-potential quasi-zero. Thus only a wall-sheath can be present at the exit of the first grid in accordance with our hypothesis.

The amplitude $\psi$ ($\delta$ in figure 1) is now fixed at 2.6 volts. Provided that ions are emitted isotropically at B, the value of Z is also known, since the square $u_0^2$ of the ion velocity at B is given by the formula for Z at amplitude $\psi = \Delta V + \delta$, used in the B-$g_2$ solution.

To determine a sensible range for X we make the reasonable assumption that in this portion of the discharge the electron current at grid 1 is essentially the cathode beam, so that the energy $v_0^2/2$ is simply the equivalent of the accelerating potential $V_{g_1}$. The latter varies between the voltage at feature **A** (~10 volts) and about 25 volts. This implies that X varies between 1.12 and 1.05.

Inserting the values of X and Z in $V(\phi)$ we find that the requirement $V(\phi)<0$ imposes a lower bound $\alpha_{min}$ on $\alpha$. Values of $\alpha$ close to $\alpha_{min}$ give rise to a quasi-zero near $\phi=2.6$ and are therefore in conflict with the stipulation that B is the *edge* of the wall sheath at $g_1$. Thus we must assume that $\alpha$ is larger than $\alpha_{min}$. Choosing $V_{g_1}$=15 volts, namely X=1.083, and using Z=13.28 (the value corresponding to $\Delta V+\delta$= 6.7) we find $\alpha_{min}$=1.4.

Figure 7c shows the wall-sheath obtained with $\alpha$=3. The distance $\Lambda$ (*now in units of the Debye length of the $g_1$-B region*) is 5.4.

In fact these solutions are not sensitive to the value of Z. Because X remains close to 1, they are also essentially independent of the accelerating potential. Moreover, adding a low-energy beam [such that $Y=(1+2\psi/w_0^2)^{1/2}$ is larger than about 6] to this part of the calculation produces insignificant changes. *This answers the crucial question italicized in the Introduction.*

Step 3: The final test

The solutions of steps 1 and 2 match smoothly at B, since on each side they culminate to a flat potential. Still, it makes no sense to claim that the primary beam dominates in the $g_1$-B region unless we know that B remains very close to the first grid.

We now show that this is indeed the case. Let $d_1$ be the distance between $g_1$ and B, $d_2$ the distance between B and $g_2$, and $d=d_1+d_2$ the total distance. Equation (15) implies that the ratio $(\Lambda/d)^2 n_0^{-1}$ is invariant. This provides a second relation between $d_1$ and $d_2$. Hence $d_1=d/(1+Rn^{1/2})$ where $R=\Lambda_2/\Lambda_1$ and $n=v_2/v_1$.

Applying this to the solutions of steps 1 and 2 we find that $d_1/d=0.1$. This completes our solution.

*8.2 The case of a larger proportion of high-potential plasma ($\beta>\alpha$)*

Obviously, the previous solution with $\beta\rightarrow\alpha$ and $X\rightarrow Z$ gives rise to the opposite configuration of largest high-potential potential plasma. The extremum of $v(\psi)$ is now a minimum and the extremum of $L(\psi)$ is a maximum. The only difference is in step 3. Because $v_{min}=1/1.961=0.51$, the ratio $d_1/d$ obtained in step 3 is twice as large (0.2). In compensation we can take a larger value of $\alpha$ in step 2 to reduce $\Lambda_1$.

*8.3 The proportion of trapped particles*

An interesting consequence of our particular choice L=1 for the Langmuir ratio is found by calculating the proportions of trapped particles [the quantities 'a' and 'b' in equations (2a) and (2b)].

Consider the case $\beta<\alpha$ of section 8.1. The proportion of trapped electrons is determined by the total charge density at $\phi=\psi$ [equation (10)]. Figure 8 shows the plot of $a(\psi)$, with an imposed value $\psi_0=6.7$ (so that $\beta=0.85/6.7$). We see that there is a range of $\psi$ surrounding $\psi_0$ where $a(\psi)$ is essentially constant [$a(\psi)\cong3$]. For $L\neq1$ this plateau *does not arise*. Instead, the function $a(\psi)$ varies up and down in this region so that varying $\psi$ in either direction affects significantly the relative proportions of trapped and drifting electrons. In other words, *the high-potential region is unstable with respect to electron-current oscillations, unless $L\cong1$*. This is in line with our earlier observations in regard to the critical behavior at feature **A** (see section 2.1) and suggests that, at optimal conditions, the system is able to settle into the $L\cong1$ solution. The fraction of trapped ions [$b(\psi)$ of equation (11)] does not exhibit this kind of fluctuating behavior. In our $L\cong1$ solution, we find $b\cong2$.

In the opposite configuration where $\beta>\alpha$, the fluctuating quantity is $b(\psi)$. Hence this case is formally unstable with respect to ion-current oscillations. Since ions are very inert, oscillations of this type are unlikely to be sustained. It follows that the relaxation oscillations observed at feature **A** are associated with the previous case involving the lighter charge-carriers (electrons) which have a much faster time of response.

In a future time dependent treatment, this difference between electron and ion dynamics will probably explain why the uphill potential rises by as much as 2.6 volts, while the analogous phenomenon (a substantial depression in the space potential below the voltage of grid 2) does not take place. A similar imbalance occurs in the field of electron and ion 'holes' [5,6].

*8.4 Physical interpretation of the threshold conditions*

The preceding results provide a reasonable qualitative explanation for the threshold requirements in regard to $\psi$ and to the cathode current.

For $\psi<4.7$, the combination of a shallow ion vortex and a deep electron vortex gives rise to a larger proportion of high-potential plasma. A significant fraction of the primary electrons undergo ionizing collisions. These share their energy with the outgoing electron and so there is a relative lack of very slow electrons ($\beta>1$). Volume ionization combined with charge-transfer collisions provides both slow and fast ions ($\alpha<<1$). These circumstances change dramatically for $\psi>4.7$. The beam of slow electrons produced on the low-potential side can now acquire sufficient energy uphill to effect excitation collisions. The resulting extra population of slower electrons generated by this feedback 'fills' the electron vortex reducing the value of $\beta$. This tends to increase the width of low-potential plasma. It also deforms somewhat the high-potential drift space so that the charge-exchange collisions conducive to creating slow ions become less probable; hence $\alpha$ increases. In the end we reach the opposite stable state of larger low-potential plasma where $\beta<<1$, $\alpha>1$.

The requirement of a sufficiently large cathode current follows from the relationship between $n_0$ and $\Lambda$ in the $g_1$-B region. Obviously the system is unable to adapt to this type of solution in the given length of intergrid distance unless $n_0$ (essentially the density of the primary beam) is large enough. In addition, if the beam is too hot, the slow-electron beam will also be wider and so the subtle conspiracy between drifting and trapped populations responsible for producing quasi-zeros can probably not take place (see section 10).

## 9. Some quantitative results

Various quantitative estimates are obtainable for testing the predictive power of the model.

Consider for instance the value of $n_0$. Using d=0.8 cm (the integrid distance in our experiment) and the result of section 8 for $\psi=6.7$ (namely $\Lambda=32$, $d_2=0.9d$) in equation (15), gives $n_0=1.1\times10^9$ cm$^{-3}$. The value for the energy $v_0^2/2$ corresponding to X is 0.06 eV. Assuming that electrons are emitted isotropically at the lower boundary we can obtain a rough estimate of the current flowing to $g_2$.

In terms of the electron energy $V_0$ in eV and the density $n_0$ in cm$^{-3}$ the current density j in Amp-cm$^{-2}$ is given by $j=9.5\times10^{-12}\times n_0 V_0^{1/2}$. Hence we find

j=2.5 mA-cm$^{-2}$. Taking an effective surface of 0.1 cm$^2$ (the area of the g$_1$ plane) gives a current of 0.25 mA. This is indeed the order of magnitude of the current measured at g$_2$ (the average anode current is two to three orders of magnitude smaller, as explained in [3]).

The primary beam current can also be estimated. Using $\Lambda$=5.4, $d_1$=0.1d, to find the density at g$_1$, and V$_0$=15 eV, yields a current of about 9 ma. This too is in agreement with what is observed: The fraction of cathode current picked up by g$_1$ at room temperature is typically about 1-1.5 ma. Because the gap between the cathode and g$_1$ is very small (0.02 cm), it is reasonable that the transmitted portion responsible for the discharge is 6 to 9 times larger.

Data on plasma parameters in experimental studies on DL formation under discharge conditions are scarce. One case that can be used to check the theory is the experiment of Leung *et al* [13]. These authors used a multiple plasma device. The DL was observed in the central 'target chamber' between two grids. The gas was argon and the amplitude $\psi$ was about 14 V. This is higher than the inelastic threshold (11.6 eV) and is consistent with our prerequisite for the case $\nu$>1. The low and high-potential plasmas were of about equal length and the values found for $\nu$ run between 1.2 and 1.5. The density n$_0$ was mentioned to be 'about' 5×10$^7$ cm$^{-3}$.

Repeating the procedure of section 8.1 with the constraints that (a): the maximum of $\nu(\psi)$ is at the imposed value, say $\psi_0$, of $\psi$, (b): $\nu(\psi_0) \cong 1.3$, (c): two quasi-zeros are formed, we find the set of parameters

$\beta\psi$=2.00  $\alpha\psi$=4.28  X=13.241  Z=16.662  $\nu_{max}$=1.34  L=0.94

The positions of the quasi-zeros of this solution are more or less symmetric so that the two plasma regions are indeed of similar length (for $\psi$=14, $\rho_{low}$ and $\rho_{high}$ are at 0.27 and 0.17 respectively).

Choosing $\psi$=14, we find $\Lambda(\psi)$=50. Assuming that d=10 cm (the distance of travel of the electron gun measuring the space potential in [13]), we find that n$_0$=3.5 ×10$^7$ cm$^{-3}$. Clearly the agreement with experiment is remarkably good.

One could argue that bringing the maxima of V($\phi$) at the quasi-zeros closer to the $\phi$ axis [namely reducing the value of $\varepsilon$ in (14)], could wildly modify these results. But there is no cause for worry. The values of $\varepsilon$ found in all our previous solutions are about $10^{-6}$. To reduce this ten-thousand-fold one must fine-tune the parameters X, Z, to 6 or 7 significant figures. Even so, the value of $\Lambda$ increases by less than a factor of two and $n_0$ by less than four. Thus the estimate of the density $n_0$ obtained for 'typically small $\varepsilon$' is a meaningful result proper to the theory.

Having said that, it is undeniable that the arbitrariness in regard to $\varepsilon$ is a weakness of the theory, and should be dealt with in a more elaborate treatment. For example, one could include in equation (5) additional stress-terms arising from energy transfer by elastic collisions and from thermal effects (a first step in this direction was taken in [7]). This might perhaps answer the question: what physical effect prevents $\varepsilon$ from becoming negative?

## 10. Individual species' densities and quasi-zero formation

The mechanism of formation of quasi-zeros can be further understood by a closer examination of the total charge density of each species.

Figure 9 shows the total electron density $n_e(\phi)$ at low $\phi$, corresponding to the solution of section 8.1 with $\psi=6.7$. We see that the characteristic variation leading to the lower quasi-zero (at $\phi \cong 0.2$) is due entirely to the *electron* density. The ions in this region merely provide a constant neutralizing background C to raise the total density towards the zero axis. C is close to 1 and is mainly made up of *trapped* ions (a 9 to 1 ratio of trapped and free ions).

This explains *'somehow number 1'* of section 2.5. When the energy of primary electrons at $g_2$ reaches the inelastic threshold (at feature **L** of figure 2), the action of the grid creates an electron vortex. Most electrons are trapped within the separatrix ($v^2=2\phi$) but some are just able to escape (small $v_0$). The combination of these electron populations gives rise to a sharp gap in the total electron distribution between the separatrix and the peak of the free electron component at $v=(v_0^2+2\phi)^{1/2}$. This gives rise to the particular variation of negative charge density shown in figure 9. A population of trapped ions (locally of uniform density) is also created by cumulative ionization. Provided that the gap is not too wide, the ions are able to neutralize the interval of smallest negative charge (the peak in figure 9). Macroscopically, a quasi-neutral, field-free scattering cell is generated where the beam can subsequently excite higher-lying states.

The same scenario with ions and electrons interchanged takes place at feature **A** and explains '*somehow number 2*'.

## 11. The alternative model with negative α and β

We end our analysis with a comparison to the model with convex distributions (truncated true Maxwellians). To avoid a sea of minus signs we change the signs of the exponents in 1(b) and 1(d) before integrating over velocities so that again $\alpha$ and $\beta$ are positive. The function $V(\phi)$ is now given by

$$V(\phi) = \frac{2\psi}{X^2-1}\left[1-\left(1+\frac{(X^2-1)\phi}{\psi}\right)^{1/2}\right] + \frac{a}{\beta}\left[2\left(\frac{\beta\phi}{\pi}\right)^{1/2} - \exp(\beta\phi)\mathrm{erf}(\beta\phi)^{1/2}\right]$$
$$+ \frac{2\nu\psi}{Z^2-1}\left[Z-\left(1+\frac{(Z^2-1)(\psi-\phi)}{\psi}\right)^{1/2}\right] + \frac{b\nu}{\alpha}\left[-2\left(\frac{\alpha\psi}{\pi}\right)^{1/2} + 2\left(\frac{\alpha(\psi-\phi)}{\pi}\right)^{1/2}\right] \quad (21)$$
$$+ \frac{b\nu}{\alpha}\left[\exp(\alpha\psi)\mathrm{erf}(\alpha\psi)^{1/2} - \exp[\alpha(\psi-\phi)]\mathrm{erf}[\alpha(\psi-\phi)]^{1/2}\right]$$

As before one eliminates a, b and $\nu$ to get $V(\phi)$ in terms of the set $\{\alpha, \beta, X, Z\}$.

The basic difference is that the beams, rather than the trapped particles, are now the dominant factor (the coefficients a and b which were previously of O(1) are extremely small here). An interesting consequence is that the values of X and Z leading to quasi-zeros can be found from power series expansions of $V(\phi)$ at the boundaries: Consider for simplicity the symmetric case $\alpha=\beta$, X=Y. Near $\phi=0$, to first order in the trapped and second order in the beam contribution, one gets: $V(\phi)=(a_1+a_2\phi^{1/2}+a_3\phi^{3/2})\phi^{3/2}$. Given the values of $\psi$ and $\alpha$, the condition that the polynomial factor (a cubic in $\phi^{1/2}$) must have a double zero gives an excellent estimate of the required value of X. The Bohm condition is now equivalent to the requirement $a_2>0$.

A solution with two non-symmetric quasi-zeros is obtained with the set $\{\beta\psi=6, \alpha\psi=3, X=2.171, Z=2.975\}$ and hence $\nu=1.07$, L=0.74. For $\psi=6.7$, the quasi zeros are at $\rho_{low}=0.05$, $\psi-\rho_{high}=0.12$.

Figure 10 shows the resulting space potential. Comparing to figure 7b, we see that now, unlike rules 1 and 2 of section 5, the highest quasi-zero is farthest from the boundary but gives rise to a narrower high-potential plasma. Also, Λ=320, one order of magnitude larger than before.

At first sight it seems that this solution too fulfills our basic objectives. Nevertheless it is incompatible with the present experimental circumstances, for several reasons:

First, the presence of the slow-electron beam responsible for displaying the inelastic spectrum is precluded by the smallness of X. Second, the value of $n_0$ (two orders of magnitude larger) is no longer in agreement with experiment. Third, ν always turns out to be very close to 1, also in conflict with observations. Fourth, it is easily verified that the physically appealing 'stability' condition based on the peak of ν(ψ) is lacking. Fifth, the conspiracy leading to quasi-zeros consists of rather miraculous fine-tuning between all four species of particles.

The difference between positive and negative trapping parameters can be further appreciated by considering an effective "temperature" $T_{tr}$ of trapped particles, obtained by averaging the energy $v^2/2$ over the distributions 1(b) and 1(d) and dividing by the density. For electrons, setting τ=βϕ, we obtain

$$T_{tr} = \phi \frac{1}{\tau} \frac{-1+e^\tau \sqrt{\tau}}{\sqrt{\pi} \text{Erfi}(\sqrt{\tau})} \qquad (22)$$

for the concave distribution, and

$$T_{tr} = \phi \frac{1}{\tau} \frac{1-e^{-\tau} \sqrt{\tau}}{\sqrt{\pi} \text{Erf}(\sqrt{\tau})} \qquad (23)$$

in the convex case.

For 0<τ<∞, $T_{tr}$ varies between (2/3)ϕ and 2ϕ in (22) and between (2/3)ϕ and zero in (23).

Charged particle temperature measurements in gas discharges are subject to the notorious 'Langmuir paradox' most recently discussed by Tsendin [18]. Nevertheless, electron temperatures quoted in the literature are usually near the threshold energy of inelastic processes. This is clearly in agreement with the range corresponding to the former case: in the solution of section 8.1, the values of $T_{tr}$ at $\phi=\psi$ corresponding to $\tau=0.85$ is $0.83\psi$. In contrast, the above solution of the convex case (where $\tau=6$) gives $T_{tr}=0.16\psi$.

## 12. Concluding remarks

I believe that the present model provides satisfactory theoretical support for the remarkable contrivance of space charge phenomena responsible for creating and maintaining the desired experimental design. Renewed experimental efforts along this line by interested readers would obviously be welcome.

In a more general context, the theory could also shed new light on the transitions from 'anode glow' to 'ball of fire' to 'Langmuir' mode in gas diodes, as defined and discussed extensively in [14-17].

An outstanding unsolved problem concerns the formation of the rising sheath adjacent to grid 1. Why does its equilibrium amplitude ($\delta$) always settle to the value of 2.6-2.7 volts? Is this number related to the kind of neutral gas?

These questions require the construction of a time-dependent model describing the explosive phenomena taking place at features **L** and **A**, which lead either to the equilibrium solution of section 8 or to current oscillations. A promising line of approach is contained in a paper by Krapchev and Ram [19]. Using a large-amplitude time-varying field modulated in space, these authors obtain a non-linear mode governed by a smooth electron distribution very similar to the jagged form employed here. An intriguing consequence is that a minimal amplitude of the potential is necessary ($\phi>2.7/T$, where T is the electron thermal energy).

## 13. Acknowledgements

I am greatly indebted to Hans Schamel whose contribution in the early stages of this project is largely responsible for the development of the present model. Martine provided badly needed encouragement and made sure of the completion of this work by creating unparalleled every-day conditions. I thank Evelyne Petermans who graciously lent me her computer, when mine broke down, so I wouldn't stop calculating quasi-zeros.

## Appendix

We wish to find the limit of $\nu(\psi)$ as $\beta\psi \to 0$, and $\alpha\psi \to \infty$.

Making the substitutions $\alpha=\gamma/\psi$ and $\beta=\delta/\psi$ in equation (12) and expanding the result to first order in $\delta$ we obtain

$$\nu = \frac{\frac{1}{\gamma} - \frac{2}{3X} + \frac{2}{1+X} - \frac{2}{\theta(\gamma)}}{\frac{1}{Z\gamma} - \frac{2}{3} + \frac{2}{1+Z} - \frac{2}{Z\theta(\gamma)}} \tag{A1}$$

where

$$\theta(\gamma) = \sqrt{\pi\gamma}\exp(-\gamma)\text{Erfi}[\sqrt{\gamma}] \tag{A2}$$

If we take a particular DL solution in this extreme range, say with $\beta\psi=6\times10^{-6}$, $\alpha\psi=60$, $\psi=6$, and seek X and Z to generate both quasi-zeros, we find that

X=8.964, Z=146, $\rho_{low}$=0.25, $\psi$-$\rho_{high}$ =0.00087.

Increasing $\alpha\psi$ raises Z but X remains very close to 9. Whence the desired limit of $\nu(\psi)$ can be found from (A1) and (A2) as $\gamma \to \infty$, $Z \to \infty$ and $X \to 9$. The function $\theta(\gamma)$ tends to 1 as $\gamma \to \infty$. Thus $\nu \to 253/90 \cong 2.81$.

# References

[01] Nicoletopoulos P 1988 Double layers as ideal grids in electron transmission spectroscopy of atoms, Proceedings of third symposium on electric double layers, ed Sanduloviciu M and Popa G (Analele stintifice ale Universitatii "Al I Cuza" din Iasi, *Serie Noua Fisica*) vol 34 suppl. pp 155-160
[02] Nicoletopoulos P 2002 *Eur. J. Phys.* **23** 533
[03] Nicoletopoulos P 2004 *Eur. J. Phys.* **25** 373
[04] Langmuir I 1929 *Phys. Rev.* **33** 954
[05] Schamel H 1982, Proceedings of first symposium on electric double layers, ed Michelsen P and Rasmussen JJ (Report R-472 Risø National Laboratory, Roskilde Denmark) pp 13-39
[06] Schamel H 1986 *Phys. Reports* **140** 161-191
[07] Schamel H, Hadjimanolaki P and Nicoletopoulos P 1990 Weak collisions in strong double layers, Physics and applications of pseudosparks, ed Gunderson MA and Schaefer G *(NATO ASI Series, Ser. B: Physics)* vol 219 (New York: Plenum) pp 277-292
[08] Raadu MA and Rasmussen JJ 1988 *Astrophys. Sp. Sci.* **144** 43-71
[09] Raadu MA 1989 *Phys. Reports* **178**, 25-97
[10] Swift DW 1975 *J. Geophys. Res.* **80**, 2096
[11] Nicoletopoulos P 2003 Analytic elastic cross sections for electron-atom scattering from generalized Fano profiles of overlapping low-energy shape resonances, *Brussels Report* (available online) unpublished
[12] Anemyia H and Nakamura Y 1984 Characteristics of double layers formed by ionization, Proceedings of second symposium on electric double layers, ed Schrittwieser R and Eder G (Inst. for theoretical physics, University of Innsbruck, A-6020 Austria) pp 188-192.
[13] Leung P, Wong AY and Quon BH 1980 *Phys. Fluids* **23** 992
[14] Malter L, Johnson EO and Webster WM 1951 *RCA Review* **12** 415
[15] Webster WM, Johnson EO and Malter L 1952 *RCA Review* **13** 163
[16] Johnson EO and Webster WM 1955 *RCA Review* **16** 82
[17] Johnson EO 1955 *RCA Review* **16** 498
[18] Tsendin LD 2003 *Plasma Sources Sci. Technol.* **12** S51-S63
[19] Krapchev VB and Ram AK 1980 *Phys Rev. A* **22** 1229

**Figure Captions**

Fig. 1 Space-potential between the cathode (K) and anode (A) showing two field-free regions joined by a double layer. The high-potential plasma is initially (solid curve) at the applied value of grid 1 voltage ($V_{g_1}$), but is raised dynamically (dashed curve) by $\delta$ volts in the course of the experiment. B indicates an appropriate cut used in the theory.

Fig. 2 Anode current plot with $\Delta V$=4.1 volts, showing a 'pure' low-potential excitation spectrum (numbered peaks to be measured from **D** at $V_{g_1}=\Delta V$). The early (lettered) peaks are excited at $V_{g_1}$ potential (the low-potential plasma has not been created as yet). Peak 1 is not relevant to the present context.

Fig. 3 The Sagdeev potential $V(\phi)$ in the symmetric case $\psi=3$, $\alpha=\beta=0.5$, X=Z, for (a) X=1.1, (b) X=6.4, (c) X=6.675.

Fig. 4 Variation of the total field-energy in the symmetric case $\psi=3$, $\alpha=\beta=0.5$, X=Z, between X=1.03 and X=6.675.

Fig. 5 The space potential in the two extreme lowest energy cases of figure 4.

Fig. 6 The variation of the total distance $\Lambda(\psi)$ corresponding to figure 4.

Fig.7a The function $\nu(\psi)$ obtained in section 8.1, for $\psi_0=5$ and $\psi_0=6.7$.

Fig.7b The space potential obtained in section 8.1, for $\psi=6.7$.

Fig.7c The sheath between grid 1 and B obtained in step 2 of section 8.1, for $\psi=\delta=2.6$.

Fig. 8 The function $a(\psi)$ obtained for $\psi_0=6.7$ from the solution of section 8.1 (with L=1.02) showing a wide plateau between about $\psi=5.5$ and $\psi=10$.

Fig. 9 The total charge density in the neighborhood of the lower quasi-zero ($\rho_{low}\cong 0.2$) obtained for $\psi=6.7$ from the solution of section 8.1 (dashed line), compared to the total *electron* density *plus one* (solid line).

Fig.10 The space potential for $\psi=6.7$, obtained using convex trapped particle distributions from the solution of section 11.

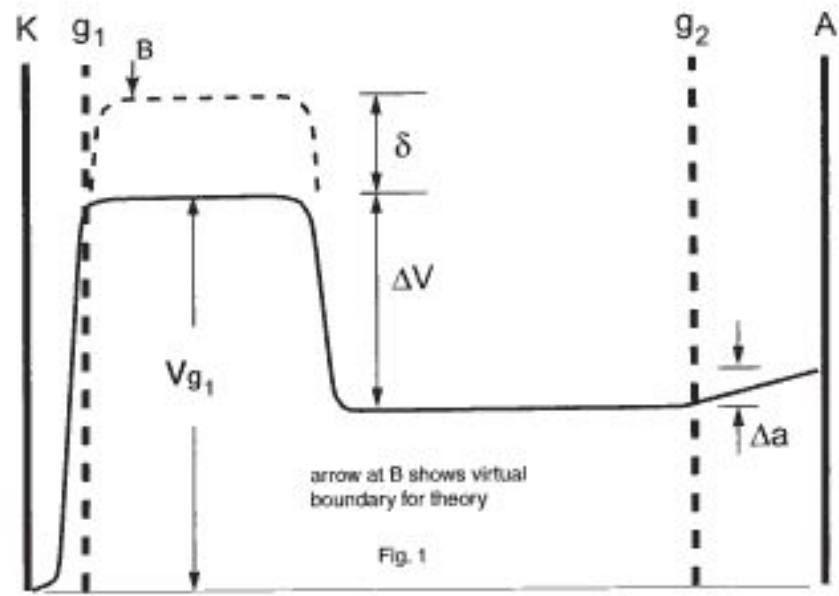

arrow at B shows virtual boundary for theory

Fig. 1

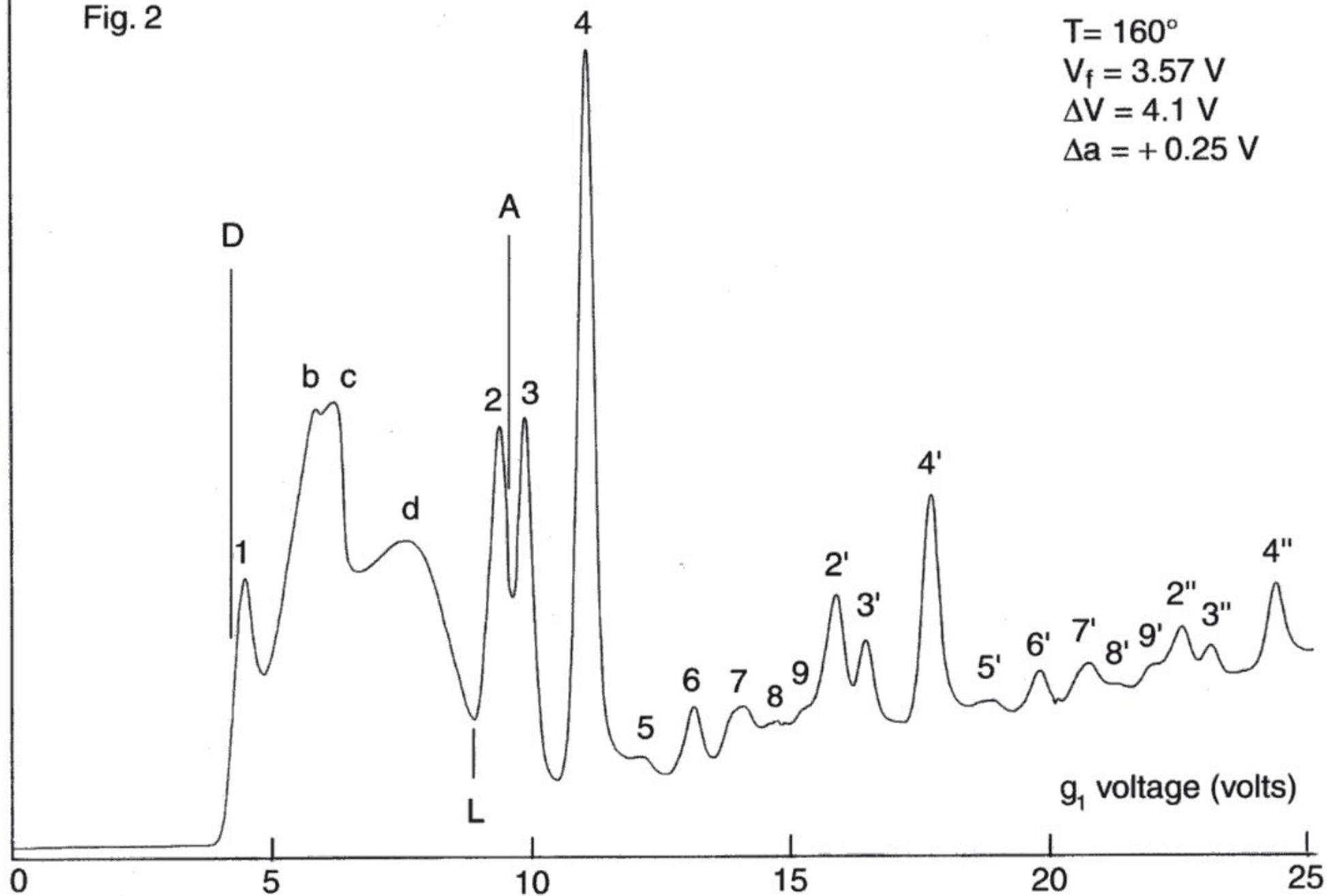

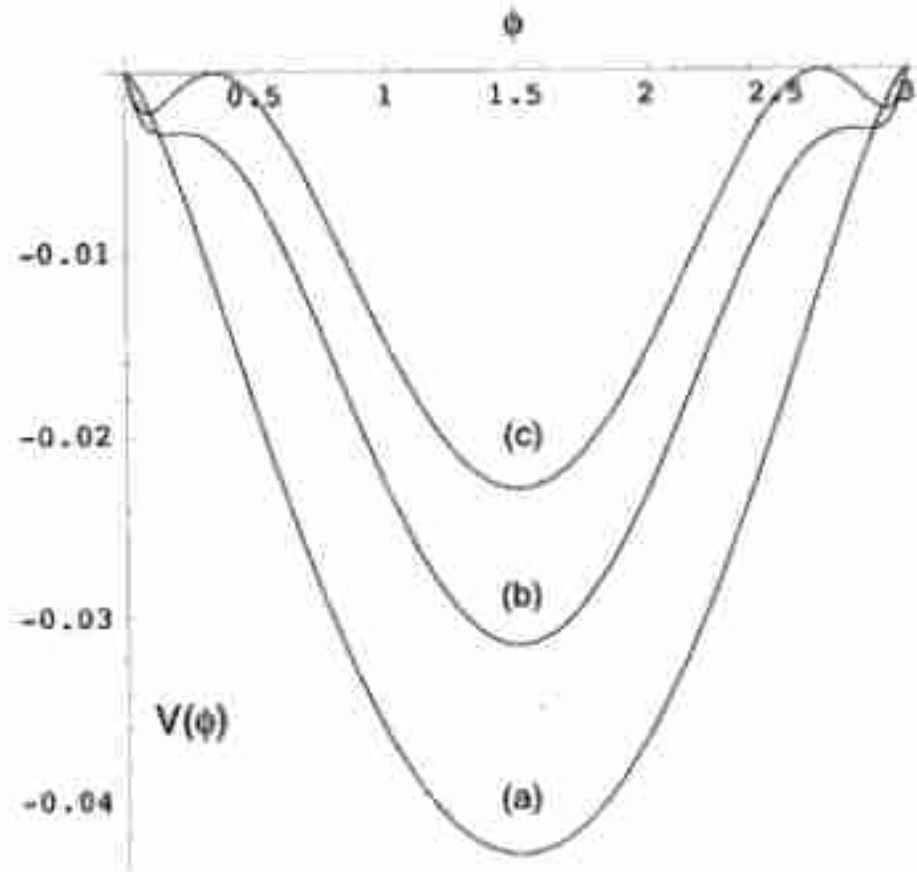

Fig. 3

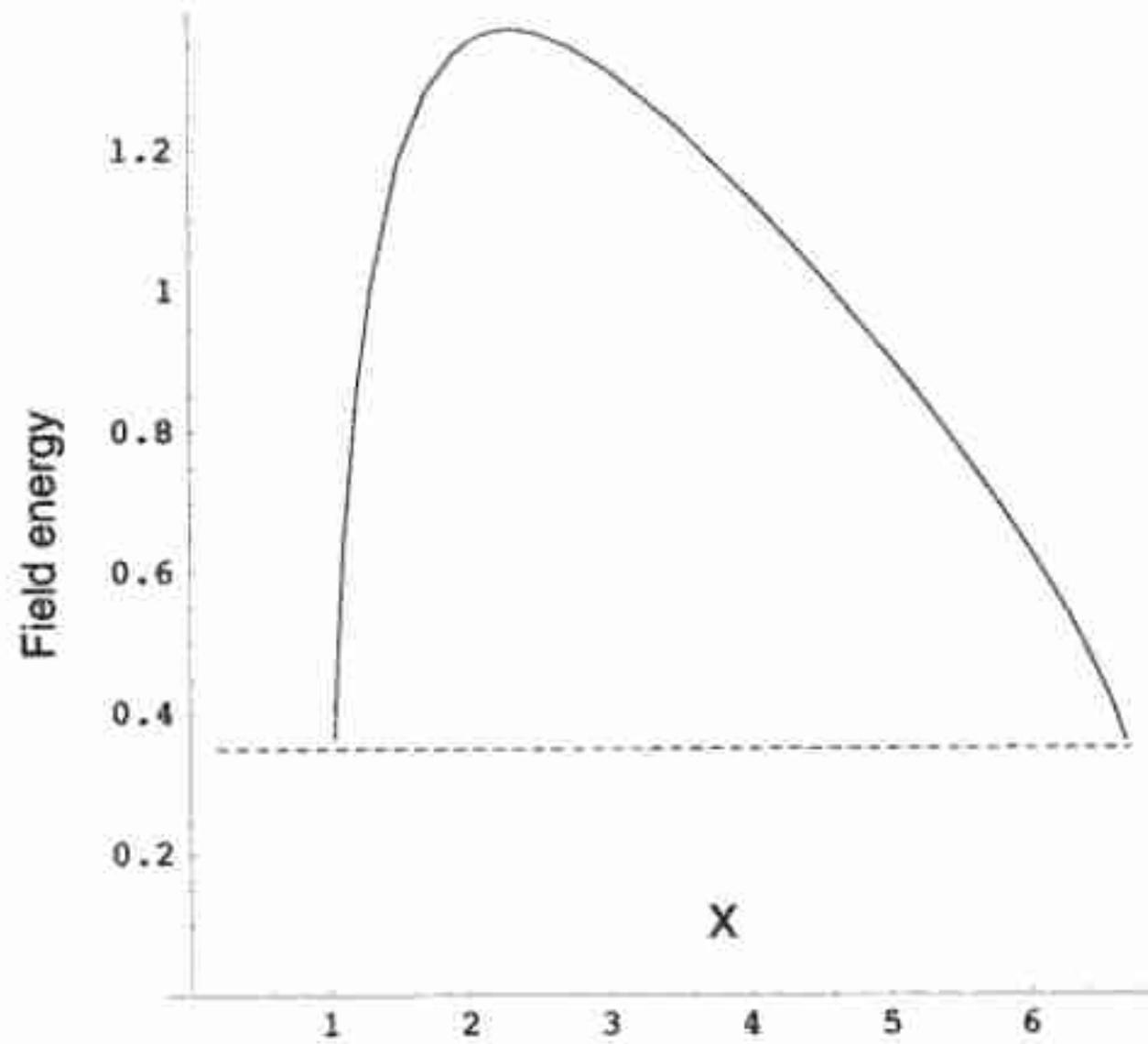

Fig. 4

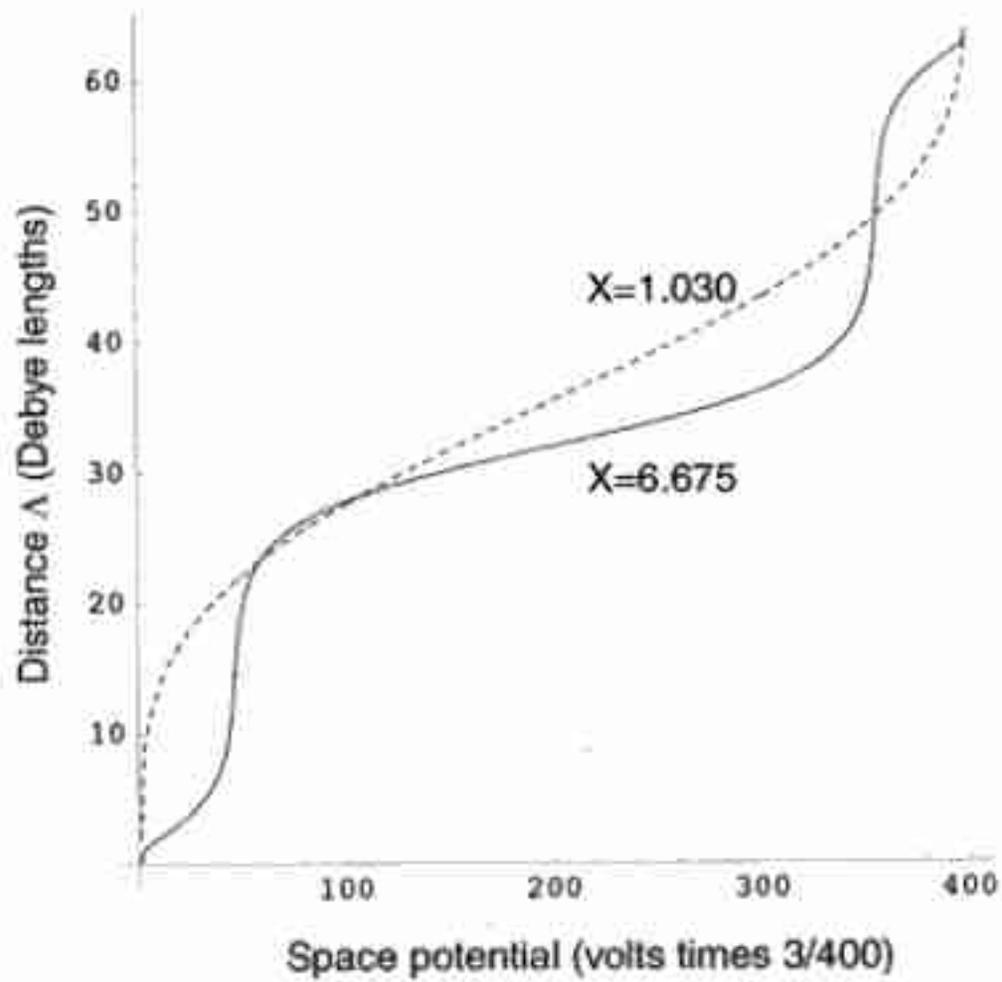

Fig. 5

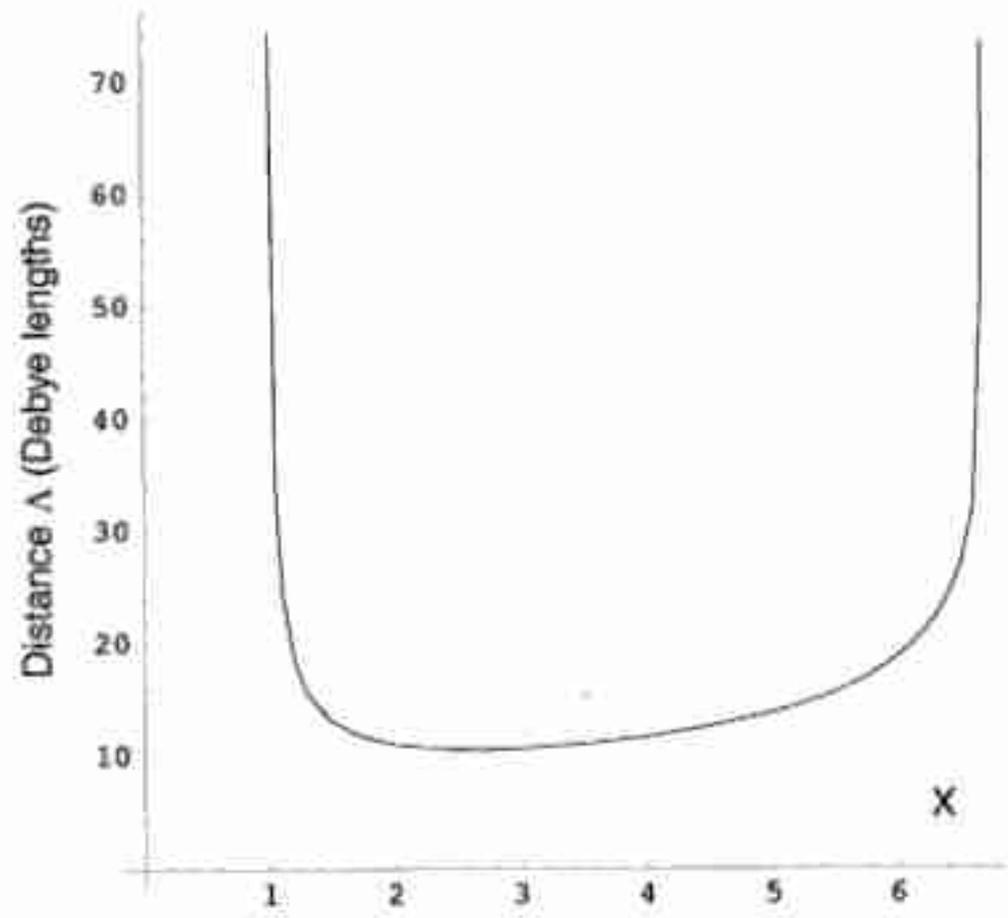

Fig.6

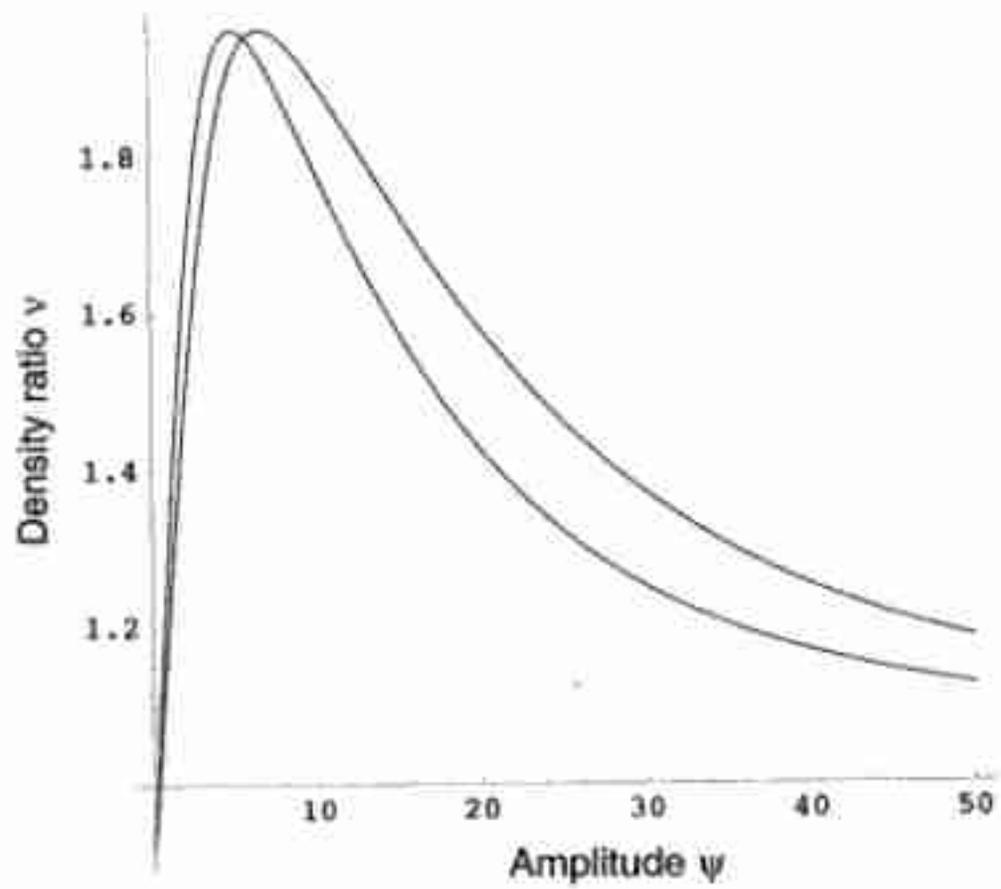

Fig. 7a

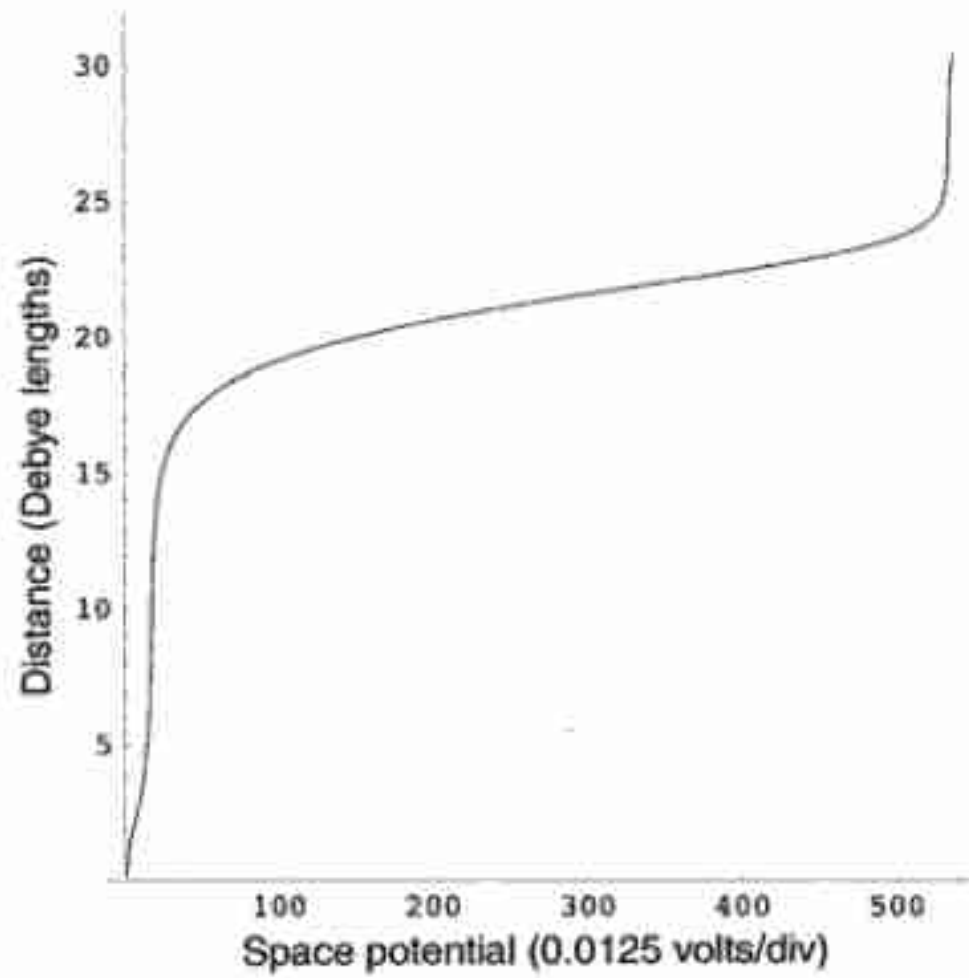

Fig. 7b

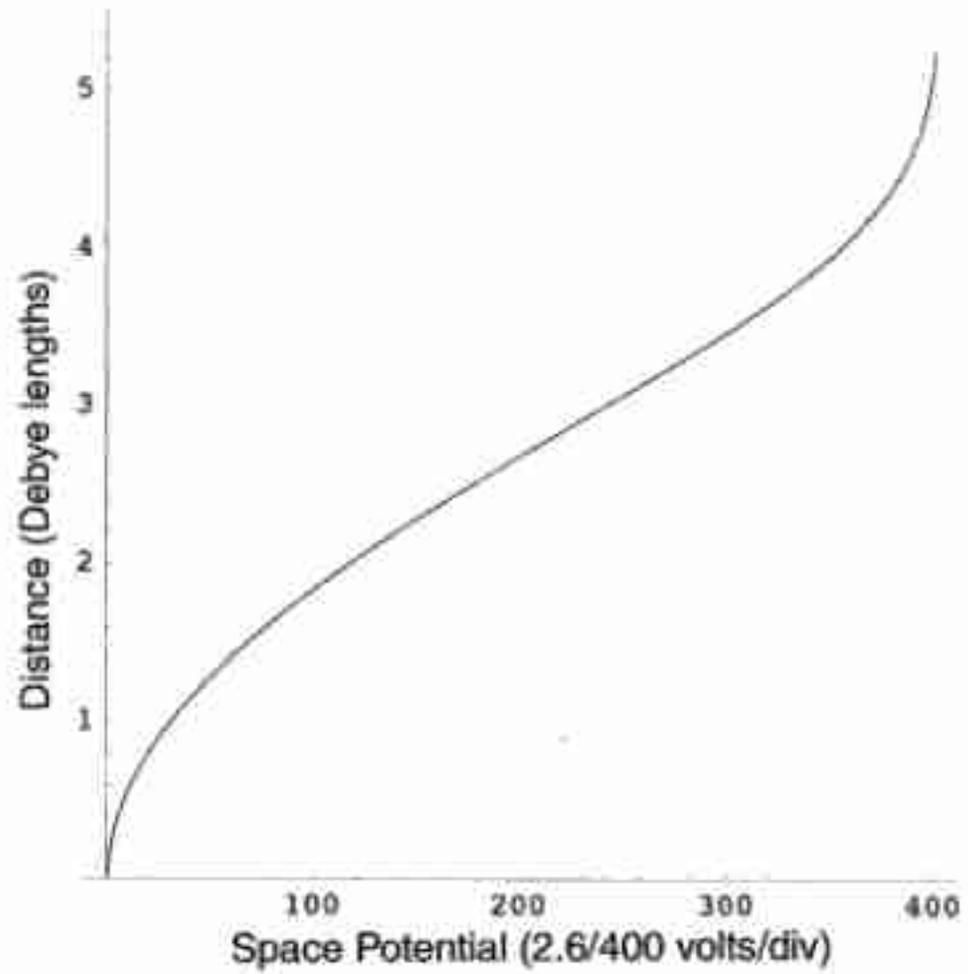

Fig. 7c

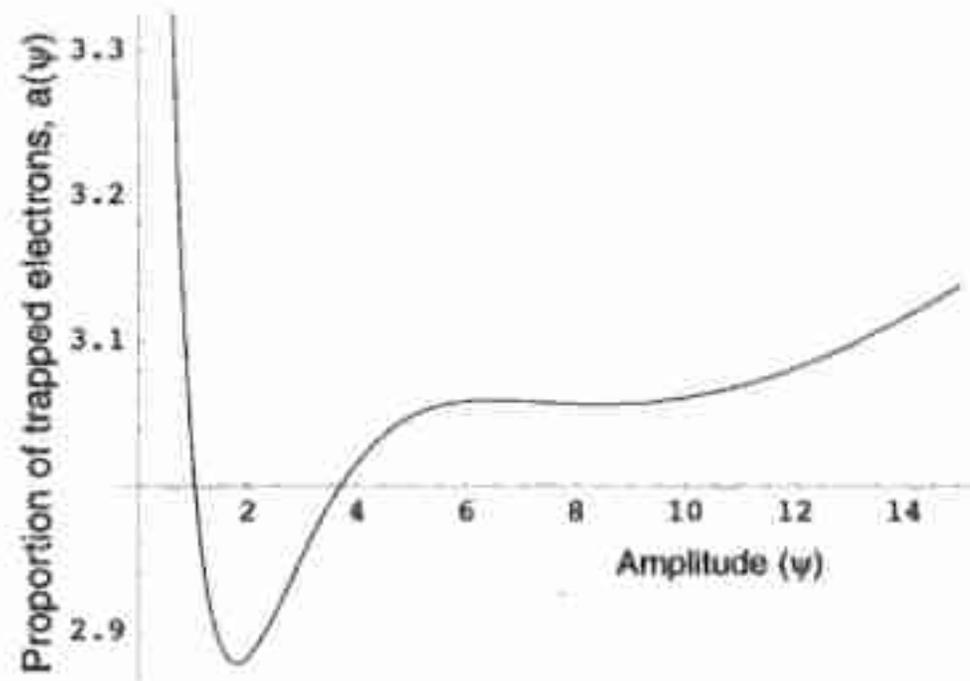

Fig. 8

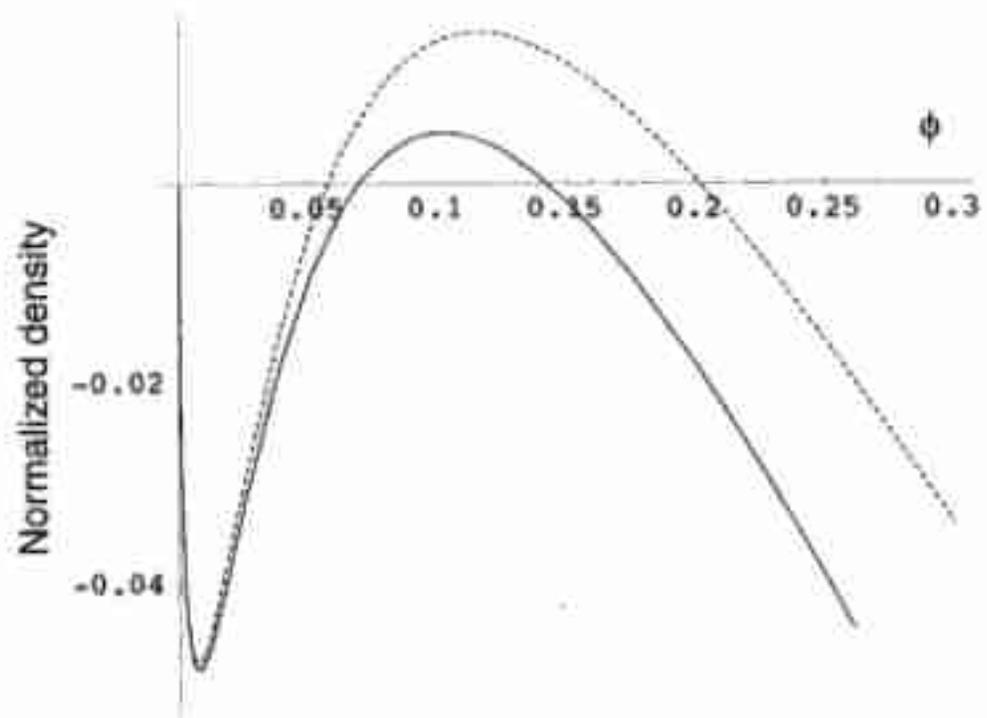

Fig. 9

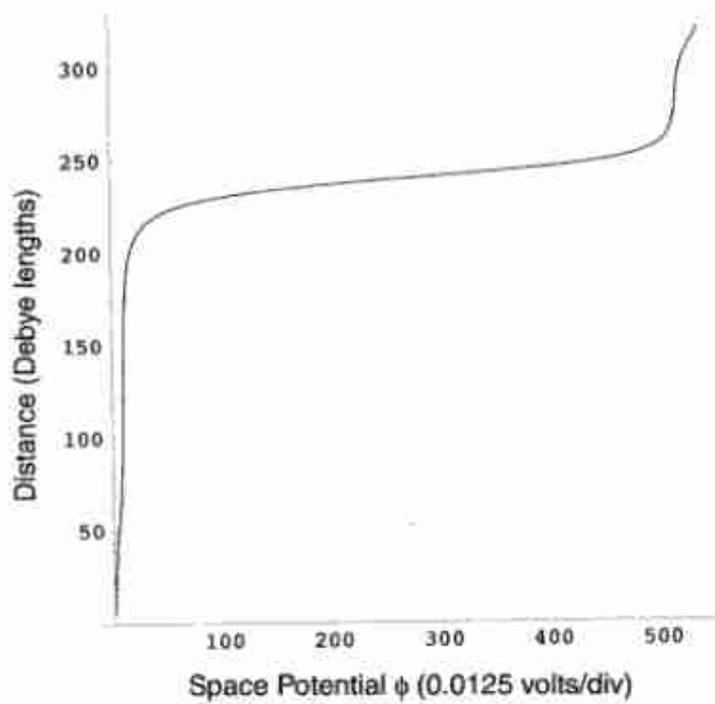

Fig. 10